\documentclass[epj]{svjour}
\usepackage{graphics}
\usepackage{appendix}
\begin{document}
\title{Cooling of a $\Lambda$-type three-level atom in a high finesse optical cavity}
\author{Lei Tan$^{1,2,}$ \thanks{Corresponding author,\emph{Present address:} tanlei@lzu.edu.cn}, Li-Wei Liu$^{1}$,   Yan-Fen Sun$^{1}$
}                     
%
%
\institute{$^{1}$Institute of Theoretical Physics, Lanzhou
University, Lanzhou 730000, China \\ $^{2}$Key Laboratory for
Magnetism and Magnetic materials of the Ministry of Education,
Lanzhou University, Lanzhou 730000,  China}
\date{Received: date / Revised version: date}
%
\abstract{ A theoretical study is carried out for the cavity
cooling of a $\Lambda$-type three level atom in a high-finesse
optical cavity with a weakly driven field.  Analytical expressions
for the friction, diffusion coefficients and the equilibrium
temperatures are obtained by using the Heisenberg equations, then
they are calculated numerically and shown graphically as a
function of controlling parameters. For a suitable choice of these
parameters, the dynamics of the cavity field interaction with the
$\Lambda$-type three-level atom introduces a sisyphus cooling
mechanism yielding lower temperatures below the Doppler limit and
allowing larger cooling rate, avoiding the problems induced by
spontaneous emission.
\PACS{32.80.Pj; 42.50.Vk.} 
} 
\maketitle
\section{Introduction}
\label{intro}

The technical developments of laser cooling and trapping greatly
promote atom manipulation, and have opened a new period in cavity
QED experimentation. New cavity cooling mechanism\cite{peter
horak,P. Maunz}, has been proposed for a single atom strongly
coupled to a high-finesse optical cavity, and the cooling
mechanism results in an extended storage time and improved
localization of atoms. An estimate shows that the observed cooling
rate is at least five times large than the cooling rate which can
be achieved by free-space cooling methods for comparable
excitation of the atom\cite{P. Maunz}. Compared to all
conventional laser cooling methods\cite{Chu,Cohen,Phillips},
cavity cooling does not rely on repeated cycles of optical pumping
and spontaneous emission of photons by the atom, but on the escape
of photons from the cavity. Moreover, as there is no requirement
in cavity cooling for a closed multilevel system, it is an
attractive approach for creating ultracold molecules\cite{M.
Viteau,J. G. Danzi}. Recently, cold atoms and molecules\cite{M. Y.
Vilensky,M. Hijlkema,T. Wilk,Ferdinand Brennecke} are now the
basis of many new areas of fundamental physics and technology, and
are the central to investigation of the Bose-Einstein
condensates\cite{J. M. Zhang,G. Szirmai}, and quantum information
processing\cite{J. H. Wu}, etc.

A single atom coupled to a single mode of an optical cavity is the
archetype model of dissipative electrodynamics. As is well known,
the atom-cavity system has two important characteristics, the
first is the cavity mirrors confining the light, which can lead to
significantly modified of atom by the cavity field. On the
contrary, the back action of the atom on the intra-cavity field is
also notable; The second is the cavity mirrors offering an extra
loss channel, which is responsible for a new dissipative and can
efficiently damp atomic motion\cite{peter horak,P. Maunz,H. J.
Metcalf,S. Nussmann}. It is well known that the cavity cooling
depends on the optical forces, as well as on the dissipation
properties. This motivates us to investigate these forces and
dissipation properties in detail. These forces are quite
substantial, which can be viewed as the dipole force and friction
force. However, dissipation inevitably leads to momentum
diffusion, which will heat the atom. In our paper, there are two
major contributions to the momentum diffusion. One is the random
momentum transfer of absorbed and emitted photons, the second
referred to the momentum diffusion is due to the fluctuations of
the dipole force. Finally, we can obtain the equilibrium
temperatures when the contributions of the friction and heating
cancel.

Even though a two-level atom model is usually used to analyze the
cavity cooling in a high-finesse optical cavity\cite{S.
Zippilli,G. Morigi,T. Salzbergur,G. Hechenblaikner,P. R. Hemmer},
the nature of a three-level atom-cavity system is quantitatively
different from that of a two-level atom system. For a there-level
atom in the optical cavity, there are more controlling parameters,
namely, two detunings, two Rabi frequencies, cavity detunings,
etc. Moreover, the extended level structure also provides
flexibility. Because of these reasons, it is of great interest to
extend the investigation on the optical forces and momentum
diffusion on the three-level atom in a high-finesse optical
cavity.  Very recently, we have studied the friction force of a
V-type three-level atoms in a high-Q cavity \cite{Liu}. Here, we
will concentrate our attention on  the variation of the friction,
diffusion coefficients and the equilibrium temperature of a
$\Lambda$-type three-level atom in high-finesse cavity, then the
full analytical expressions of the friction, diffusion
coefficients and the equilibrium temperature are obtained. For a
suitable choice of the parameters, we can obtain lower equilibrium
temperatures than the Doppler limit. For simplicity, we will only
consider one-dimensional situation.

The paper is organized as follows. In Section 2 we will give the
full analytical solutions for the optical forces and friction
coefficient by using Heisenberg equations. According to quantum
regression theorem we obtain the momentum diffusion and get the
equilibrium temperature in Section 3. Section 4 we study the
variations of the friction, diffusion coefficients and the
equilibrium temperature with detunings, coupling strengths, etc.
The friction, diffusion coefficients and the equilibrium
temperature are calculated numerically and shown graphically as a
function of controlling parameters. Some concluding remarks are
presented in Section 5.

\section{Dipole force and friction force in the $\Lambda$-type three-level atom}

\label{sec:model}

\begin{figure}
\resizebox{0.4\textwidth}{!}{%
  \includegraphics{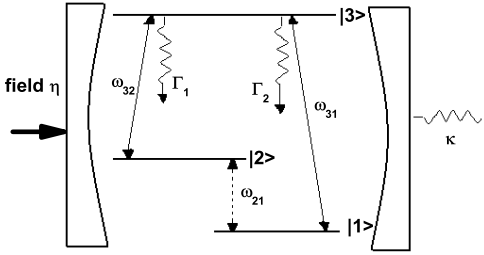}
} \caption{Configuration of a $ \Lambda $-type three-level atom
coupled to a single-mode cavity and driven by a laser field.}
\label{fig1}       
\end{figure}

Consider a $\Lambda$-type three-level atom strongly coupled to a
single mode of the electromagnetic field contained in a
high-finesse optical cavity with the cavity photons decay at a
rate $\kappa$. The schematic sketch of the basic system is given
in Fig.~\ref{fig1}. The upper level is $|3\rangle$ (energy 0),
while the two lower levels are $|2\rangle$ (energy
$-\hbar\omega_{32}$), and $|1\rangle$ (energy
$-\hbar\omega_{31}$). The transition between $|3\rangle$ and level
$|2\rangle$ ($|1\rangle$) is mediated by frequency $\omega_{32}$
($\omega_{31}$). The coupling strengths are denoted as
$g_{1}(x)=g_{0}cos(k_{1}x)$ and $g_{2}(x)=g_{0}cos(k_{2}x)$,
respectively, $k_{1}, k_{2}$ are the wave vector. The transition
between states $|2\rangle$ and $|1\rangle$ is forbidden. The state
$|3\rangle$ decays radiatively into $|1\rangle$ ($|2\rangle$) at
rate $ \Gamma_{1}$ ($ \Gamma_{2}$). In general, the system is
pumped along the cavity axis by a coherent laser filed of
frequency $\omega_{p}$ and effective amplitude $\eta$. The master
equation describing the model in Fig.~\ref{fig1} is given in the
rotating-wave approximation and in a frame rotating with the
pump frequency $\omega_{p}$ as\cite{G. Hechenblaikner}
\begin{eqnarray}
\dot{\rho}&=&-\frac{i}{\hbar}[H_{ac},\rho]
-\frac{i}{\hbar}[H_{p},\rho]+L_{\kappa}\rho+L_{\gamma}\rho,
\label{eq-1}
\end{eqnarray}
where
\begin{eqnarray}
H_{ac}&=&
-{\hbar}{\Delta_{c}}{a^{\dag}}{a}+{\hbar}{\Delta_{1}}{A_{11}}+{\hbar}{\Delta_{2}}{A_{22}}\nonumber\\&+&{\hbar}{g_{1}(x)}({a^{\dag}}{A_{13}}+{A_{31}}a)+{\hbar}{g_{2}(x)}({a^{\dag}}{A_{23}}+{A_{32}}a),\nonumber\\
H_{P} &=& {i}{\hbar}{\eta}({a^{\dag}}-{a}),\nonumber\\
{L_{\kappa}{\rho}} &=& {\kappa}{(2{a}\rho{a^{\dag}}-{a^{\dag}}{a}\rho-\rho{a^{\dag}}{a})},\nonumber\\
{L_{\gamma}}{\rho} &=&
{\Gamma_{1}}{(2{A_{13}}\rho{A_{31}}-\rho{A_{{31}}}{A_{{13}}}-{A_{{31}}}{A_{{13}}}\rho)}
\nonumber\\&+&{\Gamma_{2}}{(2{A_{23}}\rho{A_{32}}-\rho{A_{{32}}}{A_{{23}}}-{A_{{32}}}{A_{{23}}}\rho)}
\end{eqnarray}
with the atomic detunings $\Delta_{1}=\omega_{p}-\omega_{31}$ and
$\Delta_{2}=\omega_{p}-\omega_{32}$, and cavity detuning
$\Delta_{c}=\omega_{p}-\omega_{c}$. The term $H_{ac}$ represents
the energy of the atomic system and the interaction of the atom
with the cavity field, the term $H_{p}$ is the pump hamiltonian.
$L_{\kappa}\rho$ and $L_{\gamma}\rho$ describe the decay of the
resonator mode and the atomic damping to background modes,
respectively. Instead of solving the master equation one can
equivalently deal with the Heisenberg equations for atomic and
cavity filed mode operators. Considering the spontaneous emission
and cavity decay, the Heisenberg equations about the atomic
coupling to the vacuum field can be obtained\cite{Anthony E.
Siegman}.
\begin{eqnarray}
\dot{a}=i{\Delta_{c}}{a}-i{g_{1}(x)}{A_{13}}
-i{g_{2}(x)}{A_{23}}-\kappa{a}+\eta,\label{q-1-1}~\\
\dot{A}_{13}=i{\Delta_{1}}{A_{13}}+i{g_{1}(x)}(A_{33}-A_{11})a
-\Gamma_{1}{A_{13}},\label{q-1-2}\\
\dot{A}_{23}=i{\Delta_{2}}{A_{23}}+i{g_{2}(x)}(A_{33}-A_{22})a
-\Gamma_{2}{A_{23}},\label{q-1-3} \label{sss}
\end{eqnarray}
where the atomic operator $A_{ij}=|i \rangle\langle j|$ satisfies
the commutator
relation$[A_{ij},A_{kl}]=A_{il}{\delta_{jk}}-A_{kj}{\delta_{il}}$.
In order to linearize Eq. (\ref{q-1-1}) to Eq. (\ref{sss}), we
assume that the cavity is very weakly driven then the
approximation of small saturation and weak excitation of an atom
can be fitted\cite{other,other1}. In this case there is at most
one photon in the cavity and we assume that
$A_{11}$+$A_{22}$+$A_{33}$=1. For simplicity, we neglect the
population of the excited state and assume $\langle
A_{11}\rangle=F$ and $\langle A_{22}\rangle=1-F$, Thus,
\begin{eqnarray}
\langle(A_{33}-A_{11})a\rangle=-F\langle{a}\rangle,\nonumber\\
\langle(A_{33}-A_{22})a\rangle=-(1-F)\langle{a}\rangle. \label{8}
\end{eqnarray}
Then the time evolution of the expectation values can be written
as
\begin{eqnarray}
\langle\dot{a}\rangle&=&
i{\Delta_{c}}\langle{a}\rangle-i{g_{1}(x)}\langle{A_{13}}\rangle
-i{g_{2}(x)}\langle{A_{23}}\rangle-\kappa\langle{a}\rangle+\eta,\nonumber\\
\langle\dot{A}_{13}\rangle&=& i{\Delta_{1}}\langle{A_{13}}\rangle-i{F}{g_{1}(x)}\langle{a}\rangle-\Gamma_{1}\langle{A_{13}}\rangle,\nonumber\\
\langle\dot{A}_{23}\rangle &=&
i{\Delta_{2}}\langle{A_{23}}\rangle-i{(1-F)}{g_{2}(x)}\langle{a}\rangle-\Gamma_{2}\langle{A_{23}}\rangle.\label{ttt}
\end{eqnarray}

With the following definitions
\begin{eqnarray}
\mathbf{I_{\eta}}&=&\left(\begin{array}{ccc}\eta\\0\\0\end{array}\right),\nonumber\\
\mathbf{Y}&=&\left(\begin{array}{ccc}a\\A_{13}\\A_{23}\end{array}\right)\nonumber
\end{eqnarray}
\begin{eqnarray}
\mathbf{Z}&=&\left(\begin{array}{ccc}{i\Delta_{c}-\kappa}&-i{g_{1}(x)}&-i{g_{2}(x)}\\
-i{F}{g_{1}(x)}&{i\Delta_{1}-\Gamma_{1}}&0\\-i{(1-F)}{g_{2}(x)}&0&{i\Delta_{1}-\Gamma_{1}}\end{array}\right)
\end{eqnarray}
the system of linear differential equations (\ref{ttt}) can be
written in a compact matrix notation
\begin{eqnarray}
\langle\dot{Y}\rangle = \textbf{Z}\langle{Y}\rangle+I_{\eta}.
\end{eqnarray}

The steady state solution of this system of linear differential
equations is given by
\begin{eqnarray}
\langle{A_{13}}\rangle_{0}&=&
\frac{i{g_{1}(x)}F{\langle{a}\rangle}}{(i\Delta_{1}-\Gamma_{1})},\\
\langle{A_{23}}\rangle_{0}&=&
\frac{i{g_{2}(x)}(1-F){\langle{a}\rangle}}{(i\Delta_{2}-\Gamma_{2})}.\label{7}
\end{eqnarray}

Inserting Eqs. \ref{7}(a) and \ref{7}(b) into the Eq. \ref{ttt}
for the amplitude of the field we find
\begin{eqnarray}
\langle\dot{a}\rangle&=&
(i{\Delta_{c}}-\kappa-F{\gamma_{1}(x)}-i{F}U_{1}(x)-(1-F){\gamma_{2}(x)}\nonumber\\&-&i{(1-F)}U_{2}(x))\langle{a}\rangle+\eta,
\label{8}
\end{eqnarray}
where
\begin{eqnarray}
U_{1}(x) =
\Delta_{1}\frac{g_{1}^{2}(x)}{\Delta_{1}^{2}+\Gamma_{1}^{2}},
U_{2}(x) =
\Delta_{2}\frac{g_{2}^{2}(x)}{\Delta_{2}^{2}+\Gamma_{2}^{2}},\nonumber\\
\gamma_{1}(x)=\Gamma_{1}\frac{{g_{1}^{2}(x)}}{\Delta_{1}^{2}+\Gamma_{1}^{2}},
\gamma_{2}(x)=\Gamma_{2}\frac{{g_{2}^{2}(x)}}{\Delta_{2}^{2}+\Gamma_{2}^{2}},
\end{eqnarray}

The dipole force acting on the rest $\Lambda$-type three-level
atom is obtained,
\begin{eqnarray}
\textbf{F}(x) &=& \dot{\textbf{P}} =
\frac{i}{\hbar}[H,\textbf{P}]=-\hbar\nabla{g_{1}(x)}({a^{\dag}}{A_{13}}+{A_{31}}a)\nonumber\\&-&\hbar\nabla{g_{2}(x)}({a^{\dag}}{A_{23}}+{A_{32}}a)
\end{eqnarray}
and inserting Eqs.(~\ref{7},~\ref{8}) yields
\begin{eqnarray}
\langle\textbf{F}(x)\rangle_{0}=-\hbar\eta^{2}(\frac{{F}(2+F)(\Delta_{2}^{2}+\Gamma_{2}^{2}){\Delta_{1}}(\nabla{g_{1}(x)^{2}})}{|det(\textbf{Z})|^{2}}\nonumber\\+
\frac{(1-F)(3-F)(\Delta_{1}^{2}+\Gamma_{1}^{2}){\Delta_{2}}(\nabla{g_{2}(x)})^{2}}{|det(\textbf{Z})|^{2}}),
\end{eqnarray}
$det(\textbf{Z})$ is the determinant of $\textbf{Z}$, which is
given by
\begin{eqnarray}
det(\textbf{Z})=
(i\Delta_{c}-\kappa)(i\Delta_{1}-\Gamma_{1})(i\Delta_{2}-\Gamma_{2})\nonumber\\
+{F}g_{1}^{2}(x)(i\Delta_{2}-\Gamma_{2})+{(1-F)}g_{2}^{2}(x)(i\Delta_{1}-\Gamma_{1}).
\end{eqnarray}
The expression is the dipole force. It is straightforward to find
the expression for the friction force, the linear velocity
dependence of the force for small velocities ($k\upsilon<<\kappa$)
due to the cavity dynamics. Use Eq. (\ref{8}) in power of
$\upsilon$ to obtain the operator expectation values to first
order\cite{G. Hechenblaikner},
\begin{eqnarray}
\langle{a}\rangle_{1} =
\frac{\upsilon\cdot\nabla\langle{a}\rangle_{0}}{\Theta},
\label{13}
\end{eqnarray}
where
$\Theta=(i{\Delta_{c}}-\kappa-F{\gamma_{1}(x)}-i{F}U_{1}(x)-(1-F){\gamma_{2}(x)}\nonumber\\
-i{(1-F)}U_{2}(x))\langle{a}\rangle\nonumber $

 Thus the expectation value of the force operator in
first order of the velocity $\upsilon$,
\begin{eqnarray}
\langle\textbf{F}(x)\rangle_{1}&=&-\hbar(\langle{a^{\dag}}\rangle_{0}\langle{a}\rangle_{_{1}}+\langle{a^{\dag}}\rangle_{1}\langle{a}\rangle_{_{0}}
+\langle{a}\rangle_{_{1}}\langle{a^{\dag}}\rangle_{0}\nonumber\\&+&\langle{a}\rangle_{_{0}}\langle{a^{\dag}}\rangle_{1})\nabla(F(1-F)U_{1}(x)\nonumber\\&+&(1-F)(3-F)U_{2}(x))\nonumber\\
&\equiv& -\beta{\upsilon},\label{nnn}
\end{eqnarray}
$\beta$ is called friction coefficient. Inserting Equation
(\ref{13}) into an expansion of the force Equation (\ref{nnn}).
Finally one obtains the friction force in the high-finesse optical
cavity,
\begin{eqnarray}
\langle{F}\rangle_{1}&=&-\frac{8\hbar{\upsilon}\eta^{2}(\Delta_{1}^{2}+\Gamma_{1}^{2})(\Delta_{2}^{2}+\Gamma_{2}^{2})}{|det(\textbf{Z})|^{6}}\cdot\nonumber\\
\nonumber\\&+&({g_{1}(x)}(\nabla{g_{1}(x)})(F(1-F)^{2}A11\nonumber\\&+&2{F(1-F)B11+F{C11}})
\nonumber\\&+&{g_{2}(x)}(\nabla{g_{2}(x)})((1-F)F^{2}A22\nonumber\\&+&2(1-F){F}B22+(1-F)C22))\cdot\nonumber\\&+&
(F(1-F)(\Delta_{2}^{2}+\Gamma_{2}^{2})\Delta_{1}{g_{1}(x)}(\nabla{g_{1}(x)})\nonumber\\&+&(1-F)(3-F)(\Delta_{1}^{2}+\Gamma_{1}^{2})\Delta_{2}{g_{2}(x)}(\nabla{g_{2}(x)})),\label{2}
\end{eqnarray}
where
\begin{eqnarray}
A11 &=&{g_{2}^{4}(x)}(\Delta_{1}^{2}+\Gamma_{1}^{2})(2\Gamma_{2}\Delta_{1}\Delta_{2}+\Gamma_{1}(\Gamma_{2}^{2}-\Delta_{2}^{2})),\nonumber\\
A22 &=&{g_{1}^{4}(x)}(\Delta_{2}^{2}+\Gamma_{2}^{2})(2\Gamma_{1}\Delta_{1}\Delta_{2}+\Gamma_{2}(\Gamma_{1}^{2}-\Delta_{1}^{2})),\nonumber\\
B11 &=&{g_{2}^{2}(x)}(\Delta_{1}^{2}+\Gamma_{1}^{2})(\Delta_{2}^{2}+\Gamma_{2}^{2})(F{g_{1}^{2}(x)}\Gamma_{2}\nonumber\\&+&\kappa(\Delta_{1}\Delta_{2}+\Gamma_{1}\Gamma_{2})-\Delta_{c}(\Gamma_{2}\Delta_{1}-\Gamma_{1}\Delta_{2})),\nonumber\\
B22 &=&{g_{1}^{2}(x)}(\Delta_{1}^{2}+\Gamma_{1}^{2})(\Delta_{2}^{2}+\Gamma_{2}^{2})((1-F){g_{2}^{2}(x)}\Gamma_{1}\nonumber\\&+&\kappa(\Delta_{1}\Delta_{2}+\Gamma_{1}\Gamma_{2})+\Delta_{c}(\Gamma_{2}\Delta_{1}-\Gamma_{1}\Delta_{2})),\nonumber\\
C11 &=&(\Delta_{2}^{2}+\Gamma_{2}^{2})^{2}(F^{2}{g_{1}^{4}(x)}\Gamma_{1}+(2{F}\kappa{g_{1}^{2}(x)}\nonumber\\&-&2\kappa\Delta_{1}\Delta_{c}+\Gamma_{1}(\kappa^{2}-\Delta_{c}^{2}))(\Delta_{1}^{2}+\Gamma_{1}^{2})), \nonumber\\
C22 &=&
(\Delta_{1}^{2}+\Gamma_{1}^{2})^{2}((1-F)^{2}{g_{2}^{4}(x)}\Gamma_{2}+(2{(1-F)}\kappa{g_{2}^{2}(x)}\nonumber\\&-&2\kappa\Delta_{2}\Delta_{c}+\Gamma_{2}(\kappa^{2}-\Delta_{c}^{2}))(\Delta_{2}^{2}+\Gamma_{2}^{2})).~
\end{eqnarray}

The total force can be written as
\begin{eqnarray}
\langle\textbf{F}(x)\rangle\langle\textbf{F}(x)\rangle_{0}+\langle\textbf{F}(x)\rangle_{1}=\langle\textbf{F}(x)\rangle_{0}-\beta{\upsilon}
\label{eq-1}
\end{eqnarray}
The equation (\ref{eq-1}) describes the total force of the
$\Lambda$-type three-level atom within the high-finesse cavity,
the first term is dipole force, which can be found from the steady
state quantum average of the atom-cavity field interaction. The
second term is the friction force. When the friction coefficient
is positive, corresponding to cooling, while the friction
coefficient is negative, leading to heating. The value of the
friction coefficient can significantly modify the cooling process,
so it is necessary to investigate the friction coefficient.

\section{THE DIFFUSION COEFFICIENT AND THE EQUILIBRIUM TEMPERATURE}
\label{sec:temperature}

In the previous section we have obtained the friction coefficient,
when $\beta > 0$, the atom can be cooled. However, momentum
diffusion counteracts this cooling and prohibits that the atom
completely stops at rest. In our paper, there are two major
contributions to the momentum diffusion. The first referred to the
momentum diffusion is due to the fluctuations of the dipole force.
The other one is the random momentum transfer of absorbed and
emitted photons.

We first calculate the momentum diffusion coefficient due to the
fluctuations of the dipole force. The dipole force operator reads
\begin{eqnarray}
F(x)&=&-\hbar{\nabla
g_{1}(x)}{(a^{\dag}A_{13}+A_{31}a)}\nonumber\\&-&\hbar{\nabla
g_{2}(x)}{(a^{\dag}A_{23}+A_{32}a)}.\label{eq-1}
\end{eqnarray}

 For simplified calculation, we can derive equations for the
expectation values of the normally ordered operator products,
which yeilds
\begin{eqnarray}
\langle\dot{\textbf{X}}\rangle = (\Pi)_{9 \times
9}\langle{\textbf{X}}\rangle+{\eta}\langle{\Sigma}\rangle,
\label{19}
\end{eqnarray}
with $\eta$ denotes the pumping strengths of the external driving
fields and  $(\Pi)_{9 \times 9}$ is the coefficient matrix,
follows

\begin{equation}
 \Pi=
\left[
\begin{array}{ccc}
(M_{11})_{3\times 3}&(M_{12})_{3\times 3}&(M_{13})_{3\times 3} \\
(M_{21})_{3\times 3}&(M_{22})_{3\times 3}&(M_{23})_{3\times 3} \\
(M_{31})_{3\times 3}&(M_{32})_{3\times 3}&(M_{33})_{3\times 3} \\
\end{array}
\right]\,,\hspace{0.5cm}
\end{equation}
where the several  matrices $(M_{ij})_{3 \times 3}(i,j=1,2,3)$ are
given by

\begin{eqnarray}
 (M_{11})_{3\times 3}=
\left[
\begin{array}{ccc}
-(\kappa+\Gamma_{1})&0&-(\Delta_{1}-\Delta_{c}) \\
0&-(\kappa+\Gamma_{2})&0 \\
\Delta_{1}-\Delta_{c}&0&-(\kappa+\Gamma_{1}) \nonumber\\
\end{array}
\right]\,,\hspace{0.5cm}\nonumber
\end{eqnarray}

\begin{eqnarray}
 (M_{12})_{3\times 3}=
\left[
\begin{array}{ccc}
0&0&0 \\
-(\Delta_{2}-\Delta_{c})&0&0 \\
0&-2{g_{1}(x)}F&2{g_{1}(x)} \\
\end{array}
\right]\,,\hspace{0.5cm}\nonumber
\end{eqnarray}

\begin{eqnarray}
 (M_{13})_{3\times 3}=
\left[
\begin{array}{ccc}
0&0&-ig_{2}(x) \\
0&0&i{g_{1}(x)} \\
0&g_{2}(x)&0 \\
\end{array}
\right]\,,\hspace{0.5cm}\nonumber
\end{eqnarray}

\begin{eqnarray}
 (M_{21})_{3\times 3}=
\left[
\begin{array}{ccc}
0&\Delta_{2}-\Delta_{c}&0 \\
0&0&g_{1}(x) \\
0&0&-g_{1}(x)F \\
\end{array}
\right]\,,\hspace{0.5cm}\nonumber
\end{eqnarray}

\begin{eqnarray}
 (M_{22})_{3\times 3}=
\left[
\begin{array}{ccc}
-(\kappa+\Gamma_{2})&-2{g_{2}(x)}(1-F)&0 \\
g_{2}(x)&-2{\kappa}&0 \\
0&0&-2\Gamma_{1} \\
\end{array}
\right]\,,\hspace{0.5cm}\nonumber
\end{eqnarray}

\begin{eqnarray}
 (M_{23})_{3\times 3}=
\left[
\begin{array}{ccc}
2{g_{2}(x)}&g_{1}(x)&0 \\
0&0&0 \\
0&0&0 \\
\end{array}
\right]\,,\hspace{0.5cm}\nonumber
\end{eqnarray}

\begin{eqnarray}
 (M_{31})_{3\times 3}=
\left[
\begin{array}{ccc}
0&0&0 \\
0&0&-g_{2}(x)(1-F) \\
-i{g_{2}(x)(1-F)}&i{g_{1}(x)F} &0\\
\end{array}
\right]\,,\hspace{0.5cm}\nonumber
\end{eqnarray}

\begin{eqnarray}
 (M_{32})_{3\times 3}=
\left[
\begin{array}{ccc}
-g_{2}(x)(1-F)&0&0\\
-g_{1}(x)F&0&0 \\
0&0 &0\\
\end{array}
\right]\,,\hspace{0.5cm}\nonumber
\end{eqnarray}

\begin{eqnarray}
 (M_{33})_{3\times 3}=
\left[
\begin{array}{ccc}
-2\Gamma_{2}&0&0\\
0&-(\Gamma_{1}+\Gamma_{2})&-i(\Delta_{1}-\Delta_{2}) \\
0&-i(\Delta_{1}-\Delta_{2})&-(\Gamma_{1}+\Gamma_{2})\\
\end{array}
\right]\,.\hspace{0.5cm}
\end{eqnarray}

and the variable $X$ is defined as
\begin{eqnarray}
\mathbf{\textbf{X}}\equiv\left(\begin{array}{ccc}X_{1}\\X_{2}\\X_{3}\\X_{4}\\X_{5}\\X_{6}\\X_{7}\\X_{8}\\X_{9}\end{array}\right)
\equiv\left(\begin{array}{cc}a^{\dag}A_{13}+A_{31}a\\a^{\dag}A_{23}+A_{32}a\\{-i}(a^{\dag}A_{13}-A_{31}a)\\-i(a^{\dag}A_{23}-A_{32}a)\\a^{\dag}a\\A_{31}A_{13}\\A_{32}A_{23}\\A_{31}A_{23}+A_{32}A_{13}\\A_{31}A_{23}-A_{32}A_{13}\end{array}\right)
\end{eqnarray}
$\Sigma$ can be found as
\begin{eqnarray}
\mathbf{\Sigma}=\left(\begin{array}{ccccccccc}A_{13}+A_{31}\\A_{23}+A_{32}\\-i(A_{13}-A_{31})\\-i(A_{23}-A_{32})\\a+a^{\dag}\\0\\0\\0\\0\end{array}\right)
\end{eqnarray}
According to the normally ordered operators, the dipole force can
be expressed as
\begin{eqnarray}
F(x)=-\hbar{\nabla g_{1}(x)}{ X_{1}}-\hbar{\nabla
g_{2}(x)}{X_{2}}.\label{eq-1}
\end{eqnarray}

The part of the diffusion duo to the dipole fluctuation can be
written as
\begin{eqnarray}
D_{dp}&=&\hbar^{2}(\nabla{\nabla
g_{1}(x)})^{2}Re{\int_{0}}^{\infty}dt\langle\delta{X_{1}(0)}\delta{X_{1}(t)}\rangle\nonumber\\&+&\hbar^{2}(\nabla{g_{2}(x)})^{2}Re{\int_{0}}^{\infty}dt\langle\delta{X_{2}(0)}\delta{X_{2}(t)}\rangle.\label{eq-1}
\end{eqnarray}
The Eq. (\ref{19}) can be written as a homogeneous one for
$\langle{\tilde{X}}\rangle=\langle{X}\rangle-X_{0}$, where $X_{0}$
is the steady-state value of $\langle{X}\rangle$. So we get
\begin{eqnarray}
\frac{\partial\langle{\tilde{X}}\rangle}{\partial\tau}=(\Pi)_{9
\times 9}\langle{\tilde{X}}\rangle,\label{eq-1}
\end{eqnarray}
Due to the relation,
\begin{eqnarray}
\langle\delta{a_{\nu}}\delta{a_{\mu}}\rangle=\langle{a_{\nu}a_{\mu}}\rangle-\langle{a_{\nu}}\rangle\langle{a_{\mu}}\rangle,\label{eq-1}
\end{eqnarray}
We can determine the correlation the correlation function
$\langle\delta{X_{1}(0)}\delta{X_{1}(t)}$ and
$\langle\delta{X_{2}(0)}\delta{X_{2}(t)}$ from the quantum
regression theorem,
\begin{eqnarray}
\frac{\partial\langle{\tilde{X}(0){\tilde{X}(\tau)}}\rangle}{\partial\tau}=(\Pi)_{9
\times
9}\langle{\tilde{X}(0){\tilde{X}(\tau)}}\rangle,\label{eq-1}
\end{eqnarray}
 we obtain
\begin{eqnarray}
{\int_{0}}^{\infty}\langle\delta{X(0)}\delta{X(t)}\rangle{d\tau}=-(\Pi)_{9
\times
9}^{-1}\langle{\tilde{X}(0)}{\tilde{X}(0)}\rangle\equiv\Xi,\label{eq-1}
\end{eqnarray}
The diffusion coefficient can be reduced to
\begin{eqnarray}
D_{dp}=\hbar^{2}(\nabla{g_{1}(x)})^{2}\Xi_{11}+\hbar^{2}(\nabla{g_{2}(x)})^{2}\Xi_{22}.\label{eq-1}
\end{eqnarray}

The second contribution to the diffusion is the random momentum
transfer of absorbed and emitted photons, the recoil contributes to
the total diffusion by
\begin{eqnarray}
D_{s}&=&\hbar^{2}{k^{2}}\Gamma_{1}\langle{A_{31}}A_{13}\rangle+\hbar^{2}k^{2}\Gamma_{2}\langle{A_{32}A_{23}}\rangle\nonumber\\
&=&\hbar^{2}{k^{2}}\eta^{2}(\frac{g_{1}^{2}(x)F^{2}\Gamma_{1}(\Delta_{2}^{2}+\Gamma_{2}^{2})}{{|det(\textbf{Z})|}^{2}}\nonumber\\&+&\frac{g_{2}^{2}(x)(1-F)^{2}\Gamma_{2}(\Delta_{1}^{2}+\Gamma_{1}^{2})}{{|det(\textbf{Z})|}^{2}}).\label{eq-1}
\end{eqnarray}

The total diffusion can be obtained
\begin{eqnarray}
D=D_{dp}+D_{s}.\label{eq-1}
\end{eqnarray}

Hence, we can obtain the equilibrium temperature,
\begin{eqnarray}
k_{B}T=\frac{D}{\beta}.\label{eq-1}
\end{eqnarray}

Using the Heisenberg equations and the quantum regression theorem
we obtain the full analytical expressions for the friction,
diffusion coefficients and the equilibrium temperature. Because we
have many more parameters available to control the cooling process
in the $\Lambda$-type three-level atomic system, it is necessary
to study the variation of the friction, diffusion coefficients and
the equilibrium temperature with these parameters.

\begin{figure}
\resizebox{0.6\textwidth}{!}{%
  \includegraphics{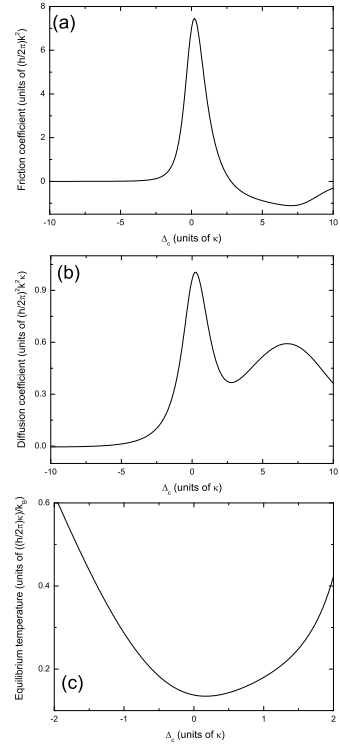}
} \caption{The average of the friction, diffusion coefficients and
the equilibrium temperature along the cavity axis over a length of
one wavelength are plotted as functions of the cavity detuning
$\Delta_{c}$. (a) the average of the friction coefficient, (b) the
average of the diffusion coefficient, (c) the equilibrium
temperature. The parameters are set to $g_{0}=8\kappa,
\Delta_{1}=8\kappa, \Delta_{2}=7\kappa, \Gamma=1.4\kappa ~and
~\eta=1.5\kappa$.} \label{fig2}
\end{figure}

\begin{figure}
\resizebox{0.5\textwidth}{!}{%
  \includegraphics{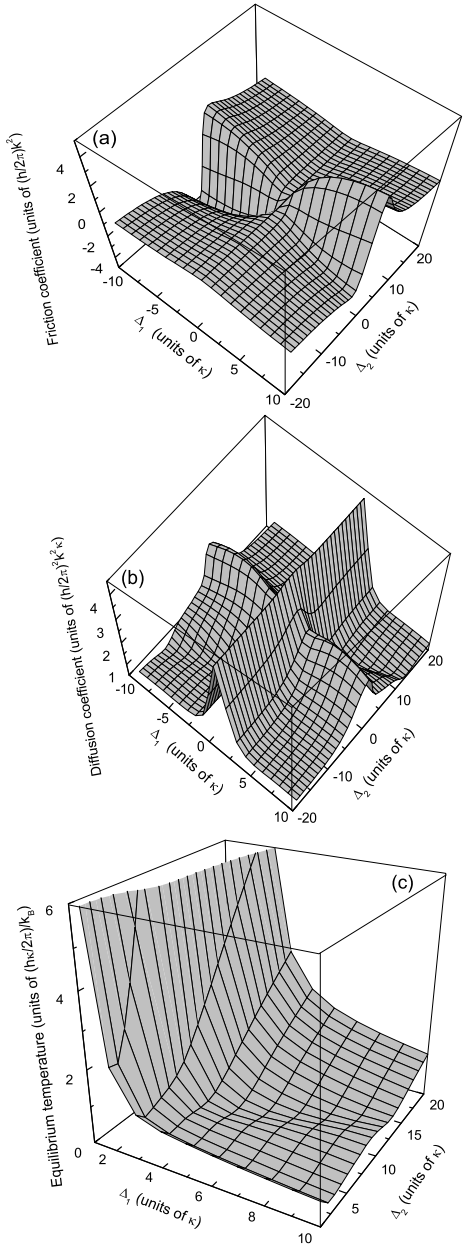}
} \caption{(a) the average of the friction coefficient, (b) the
average of the diffusion coefficient (c) the equilibrium temperature
along the cavity axis  over a length of $\lambda$ are plotted as
functions of atomic detunings $\Delta_{1}~and ~\Delta_{2}$ for
$\Delta_{c}=0$. The other parameters are set to $g_{0}=8\kappa,
\Gamma=1.4\kappa ~and ~\eta=1.5\kappa$.} \label{fig3}
\end{figure}

\section{Numerical calculation }

Using the theory described in the previous paragraphs, we have
given a detailed analysis of the cooling process of the
$\Lambda$-type three-level atom in the high-finesse optical
cavity. For simplicity, we will only consider one-dimensional
situation, and assume $K_{1}=K_{2}=k$,
$\Gamma_{1}=\Gamma_{2}=\Gamma$ in the following. Our results are
represented in scaled quantities:  the parameters
$\Delta_{c},\Delta_{1}, \Delta_{2}, g_{0}, \Gamma$ and $\eta$ are
divided by $\kappa$. The friction, diffusion coefficients and the
equilibrium temperature are calculated numerically and shown
graphically. For a suitable choice of the controllable parameters,
the equilibrium temperature can be cool down below the Doppler
limit. Fig.~\ref{fig2} shows that the average of friction,
diffusion coefficients and the equilibrium temperature along the
cavity axis over a length of one wavelength are plotted as
functions of the cavity detuning $\Delta_{c}$. The parameters are
set to $g_{0}=8\kappa, \Delta_{1}=8\kappa, \Delta_{2}=7\kappa,
\Gamma=1.4\kappa $ and $\eta=1.5\kappa$. As is well known, for
cooling a single atom in a free standing wave, the attainable
temperature is given by $T_{D}=\hbar\gamma/k_{B}$, which is the
so-called Doppler limit. Here $\gamma$ represents the nature
linewidth. In our case of cavity cooling of a $\Lambda$-type
three-level atom, the equilibrium temperature is proportion to
$\hbar\kappa/k_{B}$, $\kappa$ expresses the cavity loss rate. If
$\kappa$ is smaller than the $\gamma$, we can obtain the final
temperatures lower than the Doppler limit, which can be seen from
the Fig.~\ref{fig2}(c).

\begin{figure}
\resizebox{0.6\textwidth}{!}{%
  \includegraphics{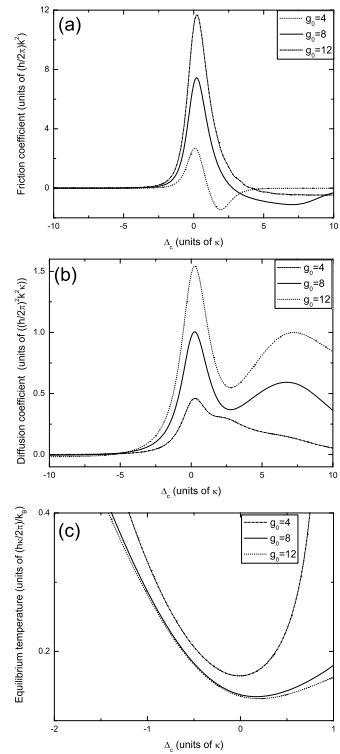}
} \caption{The dependence of (a) the average of friction, (b) the
average of diffusion coefficients and (c) the equilibrium
temperature on the coupling strengths $g_{0}=4\kappa, g_{0}=8\kappa,
g_{0}=12\kappa$. The other parameters are set to
$\Delta_{1}=8\kappa, \Delta_{2}=7\kappa, \eta=1.5\kappa,
\Gamma=1.4\kappa$.} \label{fig4}
\end{figure}

\begin{figure}
\resizebox{0.6\textwidth}{!}{%
  \includegraphics{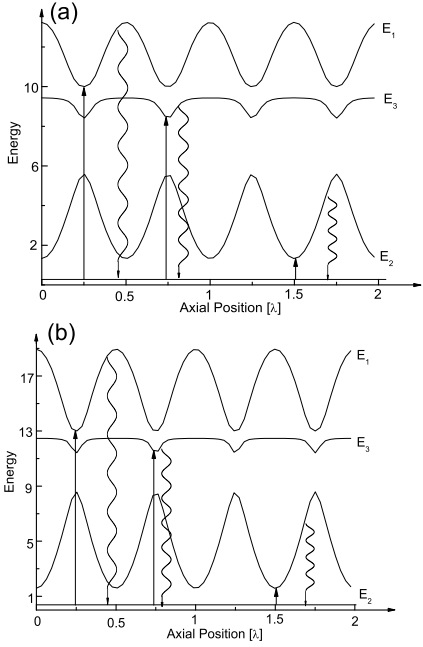}
} \caption{Energies of the $\Lambda$-type three-level atom within
cavity system (dressed states) as a function of the axial position
for (a) $g_{0}=12\kappa$, (b) $g_{0}=8\kappa$.
  } \label{fig5}
\end{figure}

\begin{figure}
\resizebox{0.6\textwidth}{!}{%
  \includegraphics{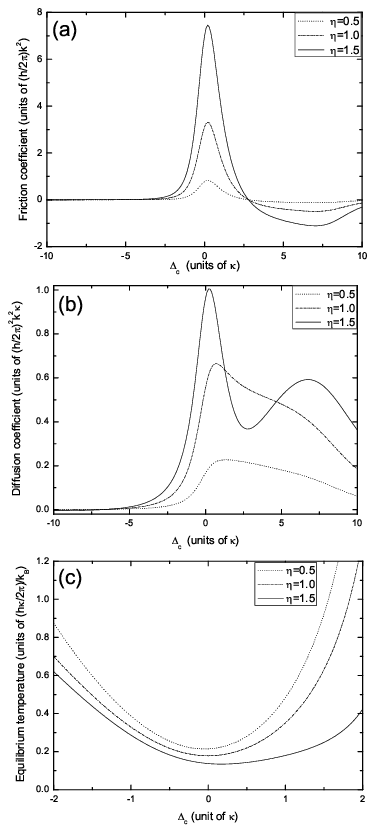}
} \caption{The dependence of (a) the average of friction, (b) the
average of diffusion coefficients and (c) the equilibrium
temperature on the pumping strengths $\eta=0.5\kappa,
\eta=1.0\kappa, \eta=1.5\kappa$. The other parameters are set to
$\Delta_{1}=8\kappa, \Delta_{2}=7\kappa, g_{0}=8\kappa,
\Gamma=1.4\kappa$. } \label{fig6}
\end{figure}

For a $\Lambda$-type three-level atom in a high-finesse cavity,
there are more parameters available to control the friction,
diffusion coefficients and the equilibrium temperature, it is
necessary to investigate the variation of the cooling process with
these parameters. To select detunings which are experimentally
advantageous for cavity cooling and trapping of the $\Lambda$-type
three-level atom, it is necessary to investigate the average of
the friction and diffusion coefficients that the atom will
experience in the high-finesse optical cavity. Figure \ref{fig3}
shows that the average of the friction and diffusion coefficients
along the cavity axis over the length of $\lambda$ are plotted as
functions of atomic detunings $\Delta_{1}~and ~\Delta_{2}$ for
$\Delta_{c}=0$. The other parameters are set to $g_{0}=8\kappa,
\Gamma=1.4\kappa $ and $\eta=1.5\kappa$. It can be found that the
atomic detunings affect the average of the friction and diffusion
coefficients significantly. For a suitable choice of the
parameters, we can obtain the low equilibrium temperature, which
can be seen from the Fig. \ref{fig3}(c). According to Fig.
\ref{fig3}, we give a especial case for a fixed values of
$\Delta_{1}$ and $\Delta_{2}$. On one hand, we consider the
dependence of the average of friction, diffusion coefficients and
equilibrium temperature on the coupling strengths, on the other
hand, we investigate the average of friction, diffusion
coefficients and equilibrium temperature with the pumping
strengths.

Firstly, we demonstrate the dependence of the average of friction,
diffusion coefficients and equilibrium temperature on the coupling
strengths $g_{0}=4\kappa, g_{0}=8\kappa, g_{0}=12\kappa$ in
Fig.~\ref{fig4}. The other parameters are set to
$\Delta_{1}=8\kappa, \Delta_{2}=7\kappa, \eta=1.5\kappa,
\Gamma=1.4\kappa$. It can found that the friction and diffusion
coefficients are enhanced, while the equilibrium temperature is
reduced with increase of the coupling strengths. As is well known,
in the strongly coupled regime, a strong friction force exists,
there is also a strong heating rate. For a batter understanding of
the strong friction force, the interpretation can use the dressed
states and sisyphus cooling mechanism. For the $\Lambda$-type
three-level atom coupled with the cavity, the eigenvalues
correspond to the energy are $E_{1}$, $E_{2}$, and $E_{3}$
respectively($\hbar\equiv 1$).
\begin{eqnarray}
E_{1}&=&\omega_{p}-\frac{(2\Delta_{2}-\Delta_{1})}{3}\nonumber\\&+&\sqrt[3]{q-\sqrt{q^{2}+p^{3}}}+\sqrt[3]{q+\sqrt{q^{2}+p^{3}}},\nonumber\\
E_{2}&=&\omega_{p}-\frac{(2\Delta_{2}-\Delta_{1})}{3}\nonumber\\&-&{\frac{1}{2}}(\sqrt[3]{q-\sqrt{q^{2}+p^{3}}}+\sqrt[3]{q+\sqrt{q^{2}+p^{3}}})
\nonumber\\&-&{\frac{\sqrt{-3}}{2}}(\sqrt[3]{q-\sqrt{q^{2}+p^{3}}}-\sqrt[3]{q+\sqrt{q^{2}+p^{3}}}),\nonumber\\
E_{3}&=&\omega_{p}-\frac{(2\Delta_{2}-\Delta_{1})}{3}\nonumber\\&-&{\frac{1}{2}}(\sqrt[3]{q-\sqrt{q^{2}+p^{3}}}+\sqrt[3]{q+\sqrt{q^{2}+p^{3}}})
\nonumber\\&+&{\frac{\sqrt{-3}}{2}}(\sqrt[3]{q-\sqrt{q^{2}+p^{3}}}-\sqrt[3]{q+\sqrt{q^{2}+p^{3}}}),
\end{eqnarray}
Where
\begin{eqnarray}
p&=&-(\frac{2\Delta_{2}-\Delta_{1}}{3})^{2}-\frac{{g_{1}^{2}(x)}+{g_{2}^{2}(x)+4\Delta_{2}(\Delta_{2}-\Delta_{1})}}{12},\nonumber\\
q&=&(\frac{2\Delta_{2}-\Delta_{1}}{3})^{3}\nonumber\\&+&\frac{(2\Delta_{2}-\Delta_{1})({g_{1}^{2}(x)}+{g_{2}^{2}(x)}+4{\Delta_{2}(\Delta_{2}-\Delta_{1}))}}{24}
\nonumber\\&+&\frac{{g_{1}^{2}(x)}(\Delta_{2}-\Delta_{1})}{8},
\end{eqnarray}
here $g_{1}(x)$ and $g_{2}(x)$ are the coupling strength,
$\Delta_{1}=\omega_{P}-\omega_{31}$ and
$\Delta_{2}=\omega_{P}-\omega_{32}$ are the atomic detunings. The
ground state energy is $E_{0}$.  Eigenvalues $E_{1}$, $E_{2}$, and
$E_{3}$ as a function of axial position are shown in Fig.
\ref{fig5}. In this atom-cavity system, cavity cooling arise when
the system is excited near a minimum of one of the dressed states,
the system will remain in this state while the atom moves further
and thus the system energy will vary according to corresponding
eigenvalues. For the $\Lambda$-type three-level atom system, after
being excited in the potential valley of the dressed state, the
stronger of the coupling strength, the steeper hills for the
$\Lambda$-type three-level atom has to climb, which can be seen in
Fig.\ref{fig5}(a) and in Fig.\ref{fig5}(b). Hence, friction
coefficient become stronger with the bigger coupling strength. On
the contrary, the momentum diffusion can be affected by the
fluctuations of atomic dipole coupled to the cavity field, because
the $\Lambda$-type three-level atom experience a stronger force
with the increase of the coupling strength, momentum diffusion
coefficient will be enhanced by increase of the coupling strength.
However, for a suitable choice of parameters, the atomic cooling
temperature can be down to lower than the Doppler temperature
limit $T_{D}$, which can be seen from the Fig. \ref{fig4}(c).

Moreover, we illustrate the dependence of the average of friction,
diffusion coefficients and equilibrium temperature on the pumping
field strengths $\eta=0.5\kappa, \eta=1.0\kappa, \eta=1.5\kappa$ in
Fig. \ref{fig6}. The other parameters are set to
$\Delta_{1}=8\kappa, \Delta_{2}=7\kappa, g_{0}=8\kappa,
\Gamma=1.4\kappa$. It can found that with the increase of the
pumping strengths, the friction and diffusion coefficients are
enhanced, while the equilibrium temperature is reduced. This is due
to the friction and diffusion coefficients are proportion to
quadratic of the pumping field, the average of friction and
diffusion coefficients can be affected remarkable by the increase of
the pumping field strength. While the equilibrium temperature is
proportion to $\hbar\kappa/k_{B}$, we can obtain the final
temperature lower than the Doppler temperature limit.

\section{CONCLUSION}
\label{sec:Conclusion}
In this paper, using the master equation and the Heisenberg
equations, we have derived the analytical solution of the dipole
force and friction force for the $\Lambda$-type three-level atom in
the high-finesse cavity. The friction coefficient can significant
affect the cooling process, however, momentum diffusion counteracts
this cooling and prohibit that the atom completely stops at rest. In
our paper, there are two major contributions to the momentum
diffusion. One is the random momentum transfer of absorbed and
emitted photons, the second referred to the momentum diffusion is
due to the fluctuations of the dipole force. According to quantum
regression theorem, we obtain the diffusion coefficient and the
equilibrium temperature. The friction, diffusion coefficients and
the equilibrium temperature are calculated numerically and shown
graphically as a function of controlling parameters. For a suitable
choice of parameters, the atomic cooling temperature can be down to
lower than the Doppler temperature limit.

\begin{acknowledgement}
 This work was partly supported by the National Natural Science
Foundation of China under Grant No. $10704031$, the National Natural
Science Foundation of China for Fostering Talents in Basic Research
under Grant No. $J0630314$, the Fundamental Research Fund for
Physics and Mathematics of Lanzhou University under Grant No.
Lzu05001, and the Nature Science Foundation of Gansu under Grant No.
$3ZS061-A25-035$.
\end{acknowledgement}

%
%

\end{document}